\documentclass[twocolumn,showpacs,preprintnumbers,amssymb,nofootinbib, aps,prd, eqsecnum]{revtex4-1}
\usepackage{graphicx, booktabs, epsf, epsfig}
\usepackage{bm} 
\usepackage{amsmath}
\usepackage{mathpazo}
\usepackage[usenames]{color}
\usepackage{hyperref}

\def\be{\begin{equation}}
\def\ee{\end{equation}}
\def\beq{\begin{eqnarray}}
\def\eeq{\end{eqnarray}}
\def \bsp{\begin{split}}
\def \ensp{ \end{split} }

\def\o{\omega}
\def\nn{\nonumber}
\def\f{\frac}

\begin{document}

\title{Tidal deformability and I-Love-Q relations for gravastars with polytropic thin shells
} 

\author{Nami Uchikata$^{1,2}$}
\email{uchikata@gwv.hep.osaka-cu.ac.jp}
\author{Shijun Yoshida$^3$}
\email{yoshida@astr.tohoku.ac.jp}
\author{Paolo Pani$^{4,5}$}
\email{paolo.pani@roma1.infn.it}
\affiliation{
$^{1}$Department of Physics, Rikkyo University, Nishi-ikebukuro, Toshima-ku, Tokyo 171-8501, Japan\\
$^2 $Department of Mathematics and Physics, Graduate School of Science, Osaka City University, Sumiyoshi-ku, Osaka 558-8585, Japan \\ 
$^3$Astronomical Institute, Tohoku University, Aramaki-Aoba, Aoba-ku, Sendai
980-8578, Japan\\
$^4$Dipartimento di Fisica, "Sapienza" Universit\`a di Roma \& Sezione INFN Roma1, Piazzale Aldo Moro 5, 00185 Roma, Italy \\
$^5$Centro Multidisciplinar de Astrof\'\i sica --- CENTRA, Departamento de F\'\i sica, Instituto Superior T\'ecnico --- IST,
Universidade de Lisboa - UL, Av. Rovisco Pais 1, 1049-001 Lisboa, Portugal}

\date{\today}

\begin{abstract}
The moment of inertia, the spin-induced quadrupole moment, and the tidal Love number of neutron-star and quark-star models are related through some relations which depend only mildly on the stellar equation of state. These ``I-Love-Q'' relations have important implications for astrophysics and gravitational-wave astronomy. An interesting problem is whether similar relations hold for other compact objects and how they approach the black-hole limit. 
To answer these questions, here we investigate the deformation properties of a large class of thin-shell gravastars, which are exotic compact objects that do not possess an event horizon nor a spacetime singularity. 
Working in a small-spin and small-tidal field expansion, we calculate the moment of inertia, the quadrupole moment, and the (quadrupolar electric) tidal Love number of gravastars with a polytropic thin shell.
The I-Love-Q relations of a thin-shell gravastar are drastically different from those of an ordinary neutron star. The Love number and quadrupole moment are negative for less compact models and the I-Love-Q relations continuously approach the black-hole limit.  
We consider a variety of polytropic equations of state for the matter shell, and find no universality in the I-Love-Q relations. 
However, we cannot deny the possibility that, similarly to the neutron-star case, an approximate universality might emerge for a limited class of equations of state.
Finally, we discuss how a measurement of the tidal deformability from the gravitational-wave detection of a compact-binary inspiral can be used to constrain exotic compact objects like gravastars.
\end{abstract}


\maketitle
\newpage

\section{Introduction}
Self-gravitating astrophysical objects are roughly spherically symmetric. If they rotate and/or are in binary systems, however, their shape deviates from spherical symmetry due to centrifugal and/or tidal forces. The response to such deformations provides us with important information on the matter distribution inside the object. 
Deviations from spherical symmetry may be characterized by the multipole 
moments of the gravitational fields outside the body. The properties of a nonspherical, self-gravitating distribution of matter are then encoded in its multipole moments. 
For instance, the spin-induced quadrupole moment encodes information about the deformability of a rotating body, which is related to the stiffness of the matter distribution (see, e.g., Refs.~\cite{Laarakkers,PoissonWill}). 

The multipole moments of a neutron star depend in similar 
ways on its internal structure. Thus, measurements of the first multipole moments
can be used to constrain the equation of state of the neutron-star core, which is still unknown at nuclear and the super-nuclear densities.

Another interesting application of the multipole moments of compact objects is related to tests of the black-hole no-hair theorems. Astrophysical black holes are fully characterized by only their mass $M$ and angular momentum $J$, all higher multipole moments being uniquely determined only by these two quantities~\cite{hage1,hage2,hage3,hage4}. Thus, independent measurements of at least three multipole moments can be used to distinguish black holes from exotic compact objects or to provide null-hypothesis tests of the Kerr metric and of general relativity (see, e.g., Ref.~\cite{Berti:2015itd} for a review). 
In this context, the lowest-order multipole moments, namely the mass $M$, angular momentum $J$, and quadrupole moment $Q$, would play a prime role.

It has been recently pointed out that there exist various approximate relations among the lowest-order multipole moments of compact stars which are only mildly dependent on the equation of state. 
By considering a reasonable set of equations of state for the neutron-star interior, approximately universal relations between the moment of inertia and the lowest-order multipole moments have been found (e.g., see Refs.~\cite{lattimer,urbanec}). 
More recently, it has been found that, for slightly deformed neutron stars, the moment of inertia $I$, the tidal quadrupole 
deformability $\lambda$ (the tidal Love number), and the spin-induced quadrupole moment $Q$, also satisfy some universal relations (dubbed as ``I-Love-Q'') that are nearly independent of the equation of state to within a few percent level~\cite{yagi, yagi2}.
Here, the tidal Love 
number $\lambda$ is defined as the ratio of the tidally-induced quadrupole moment of a nonspinning star to the strength of the perturbing quadrupolar tidal field.

These nearly universal relations will be helpful to extract physically important quantities from observational data 
if they exist within the required accuracy. For example, the I-Love-Q relations can break degeneracy in the models used for X-ray and gravitational-wave observations of neutron stars. For these reasons, nearly-universal relations of compact objects have attracted much attention in the last few years. 
Although the I-Love-Q relations were originally found~\cite{yagi,yagi2} for slightly deformed, isolated, nonmagnetized compact stars, they  have been extended to 
more general cases, namely dynamical configurations~\cite{vitor1}, rapid rotation~\cite{doneva,Pappas:2013naa,Chakrabarti:2013tca,Yagi:2014bxa}, nonbarotropic~\cite{Martinon:2014uua} and anisotropic~\cite{yagi3,yagi4,yagi5} fluids, strong magnetic fields~\cite{haskell}, and also deviations from general relativity~\cite{yagi,yagi2,Sham:2013cya,Pani:2014jra,Doneva:2014faa}. 
Possible explanations for the emergence of this approximate universality are given in Refs.~\cite{yagi7,Sham:2014kea}. 
Similar universal relations have been also found among higher-order multipole moments induced by tidal effects for nonrotating neutron stars~\cite{yagi6}. 
The inclusion of the rotational Love 
numbers in the I-Love-Q relations has been also argued in Ref.~\cite{pani3}.

These universal relations among higher-order multipole moments of a neutron star are reminiscent of the black-hole no-hair theorems, namely they show that only a handful of low-order multipole moments can (approximately) characterize the gravitational field of a compact object. 

To investigate this problem and to elucidate an origin of the I-Love-Q relation, it is interesting to understand how the I-Love-Q relations behave in the black-hole limit. Since the ratio of the mass $M$ to the radius $R$ (i.e., the compactness) of any standard spherical perfect-fluid star has the upper limit $M/R<4/9$ (in geometrized units, see, e.g., Refs.~\cite{Buchdahl:1959zz,hartle0}), the compactness of a perfect fluid star is disconnected from that of a black hole, $M/R=1/2$. To examine the black-hole limit of the I-Love-Q relations, therefore, we have to consider somewhat 
peculiar compact objects that can sustain higher compactness. Yagi and Yunes~\cite{yagi3,yagi4,yagi5} have studied this problem using fluid stars with anisotropic pressure. In the models they employed~\cite{bow}, stars with the maximally anisotropic pressure can approach the black-hole limit continuously, i.e. their compactness can be as high as $M/R\sim1/2$ in the nonrotating case. They confirmed that the universality of the I-Love-Q relations hold for stars with weakly anisotropic pressure and 
that the tidal Love number and the quadrupole moment of anisotropic fluid stars continuously approach their corresponding black hole's 
values as the compactness increases. 
On the other hand, they observed that, for strongly anisotropic pressure models, the behavior of the Love number and quadrupole moment are quite 
different from those of standard neutron stars and quark stars. These peculiar models with strongly anisotropic pressure have negative Love number and quadrupole moment, in contrast with the neutron-star case. A negative value of the spin-induced quadrupole moment basically indicates that the matter distribution of the object is prolate. 
Indeed, Yagi and Yunes observed that the rotating stars with strongly anisotropic pressure become prolate despite of the centrifugal force, due to a strongly anisotropic pressure~\cite{yagi5}.

Similar results have been also obtained for another peculiar compact star model, namely for thin-shell gravastars. The original gravastar model has been proposed 
by Mazur and Mottola~\cite{gravastar} as an alternative to the final state of the stellar evolution of very massive stars, whose collapse would form a black hole in the standard scenario. 
The thin-shell gravastar model is a simplified version of the original gravastar and is composed of a vacuum core with a positive cosmological constant (de Sitter core)  surrounded by an infinitesimally thin shell, which is required to match the interior core to the Schwarzschild exterior metric~\cite{vis}. Although this scenario lacks a precise formation mechanism, the de Sitter core is assumed to appear through a quantum phase transition in the vicinity of the would-be event horizon during the gravitational collapse of very massive objects. Gravastars are a hypothetical exotic compact objects which can be as compact as black holes but are free from the theoretical problems associated with an event horizon and a spacetime singularity~\cite{gravastar}.

The possibility of testing the gravastar scenario with electromagnetic and gravitational-wave observations has been recently argued in Refs.~\cite{BN,cr,Cardoso:2016rao,Giudice:2016zpa}. 
Some models of thin-shell gravastars are stable against small radial disturbances~\cite{vis} although nonlinear instability might 
occur for those models whose compactness is larger than $1/3$ (and therefore possess a light ring)~\cite{jk,vitor3} and for highly-spinning compact models due to the ergoregion instability~\cite{Cardoso:2007az}. 

Recently, one of us investigated the I-Love-Q relations for a particular model of thin-shell gravastar in which the energy density of the thin shell vanishes~\cite{pani2}. Similarly to the case of fluid stars with maximum anisotropic pressure, in this model, $I$, $\lambda$ and $Q$ smoothly connect to their black-hole values as the compactness increases. Furthermore, low-compactness gravastars have negative Love number and quadrupole moment and become prolate shaped when rotating. 
At the same time, by assuming that  the equation of state for the thin shell is given by the same form as that given by equilibrium sequences  of spherical solutions with fixed values of the gravitational mass, two of us~\cite{uchi2} have studied rotational effects on a thin-shell gravastar
in the absence of an external tidal field. Similarly to the case of gravastars with zero thin-shell energy, we obtained prolate shaped rotating gravastars and showed that some solutions may have the same quadrupole moment of a black hole with same mass and spin. This implies a confusion problem, namely these particular solutions cannot be distinguished from a black hole through independent measurements of their mass, spin and quadrupole moment.

Given the plethora of applications related to the multipole moments and the tidal deformability of relativistic compact objects, the motivation for the present study is manifold. On the one hand, we extend our previous work~\cite{pani2,uchi2} to investigate the universality and the black-hole limit of the I-Love-Q relations for 
various thin-shell gravastars with a generic polytropic equation of state for the shell. We use the Hartle-Thorne formalism~\cite{hartle,thorne}, 
assuming the tidal and rotational effects on the gravastar are sufficiently small, which is a standard assumption to study the tidal Love numbers and the spin-induced multipole moment for compact objects~\cite{chandra,urbanec,thorne2,flanagan,hinderer}. Since the thin-shell gravastar is composed of two distinct 
spacetimes, namely a de Sitter core and a Schwarzschild exterior in the spherically symmetric case, 
we need to take account of the junction conditions of spacetime at the thin-shell position, where the two spacetimes are matched together~\cite{israel,ba}.

On the other hand, our novel results for the tidal deformability of thin-shell gravastars allow us to investigate the extend to which gravitational-wave observations~\cite{GW150914,GW151226} can constrain gravastar models. 
Tidal effects enter the two-body inspiral gravitational waveforms at high post-Newtonian order~\cite{flanagan,hinderer} (cf., e.g., Ref.~\cite{Buonanno:2014aza} for a review). Because the tidal Love numbers of static~\cite{bin,Gurlebeck:2015xpa} (and, presumably, also rotating~\cite{pani3a,pani3}) black holes are identically zero, any gravitational-wave measurement of a nonvanishing tidal deformability would imply that one of the two objects is not a black hole. Conversely, gravitational-wave observations of compact-binary inspirals~\cite{GW150914,GW151226} may be used to put upper bounds on the tidal Love number of the two bodies, thus constraining exotic alternatives.

The plan of this paper is the following. In Sec.~II, we briefly give the formulation for constructing models of 
distorted thin-shell gravastars whose small deformation is caused by the centrifugal and the tidal forces, respectively. In Sec.~III, we present numerical results for stationary and static axisymmetric models. 
We first construct spherically symmetric, thin-shell gravastars and examine their radial stability, which depends on the equation of state for the thin-shell matter. Next, we present the results for the I-Love-Q relations of thin-shell gravastars with polytropic equation of state and investigate their black-hole limit. Finally, in Sec.~IV we discuss how a measurement of the tidal deformability from the gravitational-wave detection of a compact-binary inspiral can be used to constrain models of thin-shell gravastars.
We conclude in Sec.~V. In this study, we use geometrized units in which $G=v_c=1$, where $G$ and $v_c$ are the gravitational constant and the speed of light, respectively. 
Note that, as mentioned later, we employ the symbol ``$c$'' to denote the compactness of the object, i.e., $c:=M/R$ with $M$ and $R$ being the mass and radius of the object with spherical symmetry.   

\section{Formulation} 
Thin-shell gravastars can be constructed by matching two distinct spacetimes at the shell location. As previously discussed, we focus on small deviations from spherically symmetry, which can be investigated by using Hartle and Thorne's formalism~\cite{hartle,thorne}. 
As unperturbed solutions, we consider a spherically-symmetric thin-shell gravastar~\cite{vis} in which the de Sitter and Schwarzschild 
spacetimes are patched together. 
We are interested in the dominant quadrupole deformations and, therefore, we do not consider in detail spherically-symmetric perturbations which are also induced by rotation and by the tidal field.

As shown later, calculations of physical quantities related to the rotational and tidal deformations of the gravastar are 
very similar to each other. The differences appear only in the boundary conditions at infinity and in the time-reversal symmetry. 
Since the formulation for spin-induced deformations may be reduced to that for the tidal deformations, we first 
give a concise summary of the master equations governing slowly rotating thin-shell gravastars, as derived in Ref.~\cite{uchi2}. 
We then briefly discuss the prescription for obtaining tidally deformed nonrotating gravastars in the same framework. 
In the following, we basically follow the treatment of Ref.~\cite{uchi2} (see, also, Ref.~\cite{uchi}, in which slowly rotating regular black holes 
are studied with a similar perturbative approach).  
\subsection{Interior and exterior spacetimes of a gravastar} 
Following Hartle and Thorne's treatment~\cite{hartle,thorne}, we assume the spacetimes outside and inside the slowly rotating thin-shell gravastar to be described by a common form of the metric tensor, $g_{\alpha\beta}$, namely
\begin{align}
ds ^2&=g_{\alpha\beta}\, dx^\alpha dx^\beta \nn\\
&= -f(r)(1+2 \epsilon^2 h(r ,\theta )) dt^2 \nn\\
&\displaystyle{ + \f{1 }{f(r)}\left(1+\f{2 \epsilon^2 m(r,\theta )}{r f(r)}\right) dr^2}  \label{metric} \\
& + r^2(1+2\epsilon^2 k(r ,\theta )) \left[ d\theta ^2 +\sin^2 \theta \left\{d\phi  -\epsilon\,\o(r) dt \right\}^2 \right] \nn \\
&+{\cal O}(\epsilon^3) \nn , 
\end{align}
where $\epsilon$ stands for the smallness parameter for perturbations, and we have used 
the Greek letters ($\alpha$, $\beta$, $\gamma$, $\dots$) to denote spacetime indices.
Here, the coordinate functions $x^\alpha$ and the metric functions appearing in Eq.~\eqref{metric} are given by 
\be
\begin{split}
&(x^+)^{\mu} = (t^+,r^+,\theta^+, \phi^+), \\
& f(r)=f^ +(r^+) = 1- {2M\over r^+} ,\\
&(h(r,\theta),k(r,\theta),m(r,\theta),\o(r)) = \\
& \quad (h^+(r^+,\theta^+),k^+(r^+,\theta^+),m^+(r^+,\theta^+),\o^+(r^+)), 
\label{out}
\end{split}
\ee
for the spacetime outside the thin shell, and
\be
\begin{split}
&(x^-)^{\mu} = (t^-,r^-,\theta^-, \phi^-), \\
& f(r)=f^ -(r^-) = 1- {(r^-)^2\over L^2} ,\\
&(h(r,\theta),k(r,\theta),m(r,\theta),\o(r)) = \\
&\quad (h^-(r^-,\theta^-),k^-(r^-,\theta^-),m^-(r^-,\theta^-),\o^-(r^-)),
\end{split}
\ee
for the spacetime inside the thin shell, where 
$M$ and $L$ are, respectively, the mass of the unperturbed spherical gravastar and the de Sitter horizon radius, defined by $L=\sqrt{\Lambda/3}$, 
with $\Lambda$ being  the positive cosmological constant. Here and henceforth, the superscripts ``$+$'' and ``$-$'' indicate quantities defined 
outside and inside the thin-shell gravastar, respectively. When the distinction is not necessary, however, the superscripts are frequently omitted. 
To achieve separation of variables, we expand the perturbation functions $h(r), k(r), m(r)$  in Legendre polynomials $P_l(\cos\theta)$ 
as follows~\cite{hartle,thorne} 
\be
\begin{split}
& h(r,\theta)=h_0(r)+h_2(r)P_2(\cos\theta),\\
& m(r,\theta)=m_0(r)+m_2(r)P_2(\cos\theta),\\
& k(r,\theta)=k_2(r)P_2(\cos\theta), 
\end{split}
\ee
where equatorial symmetry is assumed and the expansion is truncated at the $l=2$ order, since we are interested only in quadrupole perturbations. 

In the absence of a perturbing tidal field, we assume that the spacetime is asymptotically flat, i.e., the metric perturbation 
must vanish in the limit of $r^+ \to \infty$.
The exterior solutions are therefore given by (see, e.g., Refs.~\cite{hartle,thorne}) 
\begin{align}
&\o ^+  = \f {2 J} {r^3} , \label{rotate1} \\
& m_0^+ =\delta M -\f{J^2 }{r^3}, \\
& h_0 ^+= - \f{\delta M}{r- 2M } + \f{J^2 }{r^3 (r- 2M)}  ,\\
& h _2 ^+= J^2 \left (\f{1}{ M r^3} + \f{1}{r^4}\right ) +  B Q_2 ^{\, 2} \left (\f{r}{M} -1 \right ) ,\\ 
& k _2 ^+=- \f{J^2}{  r^4} -  B\f{2 M}{ \sqrt {r (r-2 M)}} Q_2 ^{\, 1} \left (\f{r}{M} -1 \right )-h_2^+ ,\\
& m_2  ^+= (r-2 M)  \left (- h_2 ^+ +\f{r^4}{6}\left (\f{d\o^+}{dr} \right ) ^2 \right ), 
\label{rotate2}
\end{align}
where $J$ and $\delta M$ are the angular momentum and the spin-induced mass shift, respectively.
Here, $B$ is an integration constant, through which the quadrupole moment of the rotating gravastar is defined as
\be
 Q = {J^2\over M}+{8\over 5}\, BM^3 \, . \label{def_Q}
\ee
Note that, for slowly rotating Kerr black holes, regularity at the horizon imposes $B=0$.
The function $Q^m_l$ is the Legendre function of the second kind. The explicit forms of $Q_2^{\, 2}$ and $Q_2^{\, 1}$ read
\begin{align}
Q_2^{\, 2}(x) & = \f{x(5-3 x^2)} {x^2 -1} +\f{3(x^2 -1)}{2}  \log \f {x+1} {x-1},\\
Q_2^{\, 1}(x)& = \sqrt{x^2 -1 } \left (\f{2-3x^2}{x^2-1}+ \f {3 x}{2} \log \f {x+1} {x-1} \right ) .
\end{align}
The interior solutions that are regular at the center of the star are given by (see, e.g., Refs.~\cite{uchi,uchi2,pani2}) 
\begin{align}
& \o ^ -= C_1, \label{rotate3}\\ 
& m_0 ^ -=0, \\
& h_0 ^-=C_2 ,\\
& h _2^- = \f { C_3}{8 r^2} \left (\f {-3 L^2 + 5  r^2} { L^2  f^ - (r) } + \f {3 L f^-(r) \mbox{Arctanh} (r/L)}{r} \right ), \label{h2n}\\ 
& k _2 ^-=\f { C_3}{8 r^2 L} \left (\f {3 L^2 + 4  r^2} { L } - \f {3 (L^2+r^2) \mbox{Arctanh} (r/L)}{r} \right ), \label{k2n} \\
& m_2  ^-= -r f^ -(r) h_2 ^-, \label{rotate4}
\end{align}
where $C_1$, $C_2$ and $C_3$ are integration constants.
The exterior and interior perturbed metric are matched so as to fulfill the junction conditions, as we discuss in the next section.

Now let us turn our attention to the case of the tidal deformations of a spherically symmetric gravastar. We focus on gravastars whose deformation is induced by a static and axially symmetric
\footnote{In the case of a nonspinning, tidally deformed object, the spherical symmetry of the background configuration guarantees that perturbations with different azimuthal number $m$ decouple from each other and are degenerate (i.e., the radial functions are independent of $m$). Therefore, we can consider axial symmetry and set $m=0$ without loss of generality. After the radial functions are obtained, it is straightforward to derive the full, nonaxisymmetric, deformed metric.}  external tidal field. Thus, we have to assume 
(i) $\omega=0$ and (ii) the spacetime is not asymptotically flat because of the existence of a tidal source on the symmetry axis at a sufficiently large distance. 
These requirements change the solutions given in Eqs.~\eqref{rotate1}--\eqref{rotate2} and \eqref{rotate3}--\eqref{rotate4} as follows:  
\begin{align}
J&=0 \, , \\
 h _2 ^+&= -D_1 Q_2 ^{\, 2} \left (\f{r}{M} -1 \right )-3D_2 \f{r^2}{M^2} \left (1-\f{2M}{r}  \right )  , \label{h2t} \\ 
 k_2^+ & =  D_1\f{2 M}{ \sqrt {r (r-2 M)}} Q_2 ^{\, 1} \left (\f{r}{M} -1 \right ) \nn\\
 & \quad +6D_2 \left(\f{r}{M}-1\right )-h_2^+  , \\
 m_2^+ &=-(r-2M) h_2^+, \\
 C_1&=0 \, , 
 \end{align}
where $D_1$ and $D_2$ are new integration constants. The other perturbation functions remain invariant. 
The asymptotic behavior of $h _2 ^+(r)$ as $r \to \infty$ reads 
\be
\begin{split}
h _2 ^ + \to &  -\f{8}{5}\left (\f{M}{r} \right )^3\,\left\{ 1 + O\left ( \f{M}{r} \right )  \right\}D_1 \\
&-3\left (\f{r}{M} \right )^2\,\left\{ 1 + O \left ( \f{M}{r} \right )\right\} D_2 \,.
\label{love0}
\end{split}
\ee
Since $-(1+g_{tt})/2$ at large distance from an isolated star may be regarded as the Newtonian gravitational potential~\cite{hinderer}, 
the first and second lines in the right-hand side of Eq.~\eqref{love0} may be interpreted as the induced quadrupole component of 
the gravitational potential and the quadrupole external tidal potential, respectively. 
The tidal Love number, $\lambda$, is then defined by (see, e.g.,~\cite{hinderer,PoissonWill}) 
\be
\lambda =  \f{8}{45}\f{D_1}{D_2}M^5.
\label{love}
\ee

\subsection{Junction conditions for gluing the two spacetimes} 
To match the two spacetimes in a physically appropriate way at the location of the thin shell, we need to impose the so-called 
junction conditions~\cite{israel,ba}. 

The location of the thin shell is given by 
\be
(x^\pm)^{\mu} = (x^\pm)^{\mu}(y^a) , 
\ee
where $y^a$ are the intrinsic coordinate functions of the thin shell. Here and henceforth, the roman letter indices  ($a$, $b$, $c$, $\dots$) are used to indicate the tensor quantities defined on the three-dimensional hypersurface of the thin shell. 
We denote the intrinsic coordinate functions as $y^a=(T, \Theta, \Phi)$ and define the thin-shell location as 
\begin{align}
(x^{\pm})^{\mu} &= (A^{\pm} T, R + \epsilon ^2\xi^{\pm}(\Theta), \Theta+\epsilon^2 (l^{\pm})^{\Theta}(\Theta), \Phi) \nn \\
&+{\cal O}(\epsilon^3), 
\label{shell-coord}
\end{align}
where $A^{\pm}$ and $R$ are time rescalings and the radius of the gravastar for unperturbed spherical states, respectively, 
and $\xi^{\pm}$ and $(l^{\pm})^{\Theta}$ stand for functions of $\Theta$ related to the displacement of the thin-shell position.  
The radius of spherically symmetric gravastars $R$ satisfies $L>R$ and $2M<R$ because the gravastars possess no horizon.  
By using the degree of freedom of the coordinate choice~\cite{uchi2}, we may assume, without loss of generality, that $A^+ = 1$ and $(l^+)^{\Theta}=0$.
To achieve separation of variables, we also decompose the displacement in Legendre polynomials,  
\be
\xi (\Theta) = \xi_0 + \xi_2 P_2(\cos \Theta). 
\ee
A set of three independent tangent vectors to the thin shell, 
$e^{\mu} _a$, is given by 
\be
e^{\mu} _a =\f {\partial x^{\mu}} {\partial y^a}. 
\ee
The unit normal vector to the thin shell, $n^\mu$, is another primary quantity characterizing the thin shell.

The first junction condition shows how the spacetimes induced on the thin shell are matched smoothly in an intrinsic sense.  
It states that the induced metric $h_{ab}:=g _{\mu \nu} \, e^{\mu} _a e ^{\nu} _b $ is continuous through the thin shell, namely
\be
 [[h_{ab}]] =0\,, 
\label{junc1}
\ee
where the double square brackets indicate
\be
[[E]] =E^+ - E^-\,,
\ee
for the generic quantities $E^{\pm}$ defined on the thin shell.

From the first junction condition, Eq.~\eqref{junc1}, and from the thin-shell location, Eq.~\eqref{shell-coord}, we get 
the following matching conditions (for details, see Ref.~\cite{uchi2});
\begin{align}
&(A^-)^2 = \f {f^+(R)} {f^-(R)} , \quad  \o ^- =C_1 = \f {2J} {R^3A^-} , \label{match1} \\
&[[\xi _ 0 ]]= 0 \,, \quad (l^-)^{\Theta}=0 \,, \\ 
&  [[h_0 (R)]] +\f {R \xi^-_0} {L^2 f^-(R)} + \f {M \xi^+_0} {R^2 f^+ (R)} =0, \\
& \left [\left [\f { \xi_2} {R} + k_2  (R)\right ]\right ]= 0,  \label{match2} \\ 
& [[h_2 (R)]] +\f {R \xi^-_2} {L^2 f^-(R)} + \f {M \xi^+_2} {R^2 f^+ (R)} =0. \label{match3}
\end{align}

The second junction condition determines the stress energy tensor $S_{ab}$ on the thin shell, 
\be
S_{ab} = {1\over 8\pi}\left([[ K_{ab} ]] -h_{ab} [[K]] \right),
\label{sab}
\ee
in terms of the extrinsic curvature $K_{ab}$, which is defined in terms of the tangent vectors $e^{\mu}_a$ and of the unit normal vector $n^{\mu}$ 
to the thin shell as 
\be
K_{ab} \equiv  - n _{\alpha ; \beta} e^{\alpha} _a e ^{\beta} _b \quad \mbox{and} \quad  K=h_{ab} K^{ab}\,.
\ee
Here, the semicolon $(;)$ denotes the covariant derivative associated with the metric function $g_{\alpha\beta}$. 
The nonzero components of the normal vector to the thin shell up to an accuracy of $\epsilon^2$ are given by
\begin{align}
n_r & =\f{1}{\sqrt{f}}+\epsilon^2 \f{m_0+m_2 P_2(\cos \theta)}{r {f}^{3\over 2}} +{\cal O}(\epsilon ^4),\\
n_{\theta} & = -\epsilon ^2 \f{\xi_2}{\sqrt{f}} \partial_\theta P_2 (\cos \theta)+{\cal O}(\epsilon ^4).
\end{align}
The nonzero components of the extrinsic curvature up to an accuracy of $\epsilon$ are then given by 
\begin{align}
& K^T_T= -\f{f^{\prime}}{2\sqrt{f}} + {\cal O}(\epsilon^2) , \\
& K^{\Theta}_{\Theta}=K^{\Phi}_{\Phi}= - \f{\sqrt{f}} {R} + {\cal O}(\epsilon^2), \\
& K^{\Phi}_T =\epsilon \f{ f (2 \o + R \o ^{\prime})-R \o f^{\prime} }{2 R \sqrt{f} } A+ {\cal O}(\epsilon^3) ,\\
& K^T_{\Phi} =-\epsilon \f{R^2 \o^{\prime}}{2 A \sqrt{f} } \sin^2 \Theta + {\cal O}(\epsilon^3) , \\
& K =- \f{4f+Rf^{\prime}}{2R\sqrt{f}} + {\cal O}(\epsilon^2). 
\end{align} 
The $\epsilon^2$-order extrinsic curvature is lengthy and is given in Appendix A. To specify 
the second junction conditions concretely, as argued later, we need to prescribe some equation of state for the thin-shell matter. 

\subsection{Matter properties of the thin shell: Perfect-fluid thin shell} 
The energy density $\sigma $ of the thin shell is defined by the eigenvalue of the stress-energy tensor $S_{ab}$,
\be
S^a_{b} u^b=-\sigma u^a,
\label{eigen}
\ee
where $u^a $ is the matter velocity tangent to the thin shell satisfying $u_a u^a = -1$.
By using the projection tensor $q_{ab}= h_{ab}+u_a u_b$, we may define the projected stress tensor of the thin shell, namely
\be
\gamma_{ab} = S^{cd} q_{ac}q_{bd}\,. 
\ee
We consider a perfect-fluid thin shell, whose stress-energy tensor reads $\gamma_{ab} = p q_{ab}$ with $p$ being the isotropic pressure of the fluid thin shell. We also expand the energy density and the pressure 
of the thin shell as follows: 
\be
\begin{split}
\sigma&=\sigma_0+\epsilon^2 \delta \sigma +{\cal O}(\epsilon^4)  \\
&\equiv\sigma_0+ \epsilon^2(\delta \sigma_0 + \delta \sigma_2 P_2)+{\cal O}(\epsilon^4)  \,, \\
p&=p_0+\epsilon^2 \delta p +{\cal O}(\epsilon^4) \\
&\equiv p_0+ \epsilon^2(\delta p_0 + \delta p_2 P_2)+{\cal O}(\epsilon^4)  \,,
\end{split}
\ee
where $\sigma_0$ and $p_0$ are the energy density and pressure for the spherical gravastar, respectively, 
and $\delta\sigma$ and $\delta p$ are the energy density and pressure perturbations about the spherical gravastar, respectively. 
The latter perturbations are also expanded in Legendre polynomials $P_2\equiv P_2(\cos\theta)$. By combining the perfect-fluid condition with the first junction condition, summarized in Eqs.~\eqref{match1}--\eqref{match3}, and 
by specifying the equation of state for perturbations $\delta p=\delta p(\delta \sigma)=(dp/d\sigma)\,\delta\sigma$, as argued later, 
we can obtain the quadrupolar deformations of slowly rotating and tidally deformed gravastars. 

\subsubsection{Master equations for spherically symmetric unperturbed solutions} 
In the unperturbed state, the only nonzero component of $u^a$ is $u^T = 1/\sqrt{f}$, then we get
\beq
\sigma_0&=  & \f{\sqrt{f^-}-\sqrt{f^+}}{4 \pi R}, \label{def_sig0} \\
p_0 &= & \f{1}{8 \pi R^2} \left ( \f{R-M}{\sqrt{f^+}} -R\f {1-2R^2/L^2}{\sqrt{f^-}}\right ). 
\label{def_p0}
\eeq
Note that the pressure of the thin shell in the unperturbed state is positive for any positive values of $R$ and $M$ provided  
$f^+>0$ and $f^->0$.

\subsubsection{Master equations for the $\epsilon$-order solutions } 
In the case of a slowly rotating gravastar, we assume the thin shell to be uniformly rotating, i.e., 
the angular velocity $\Omega$ of the shell is constant.
Then, the components of $u^a$ satisfies  
\be
{u^{\Phi} \over  u^T}={d\Phi\over dT}=\Omega\,, \quad u^\Theta=0\,. \label{SeigenP}
\ee
We further assume that $\epsilon \equiv   \Omega / \Omega_k \ll 1$, where 
$\Omega_k = \sqrt{M/R^3}$ is the Keplerian frequency of the spherical gravastar. 
From the $T$ and $\Phi$ components of Eq. \eqref{eigen}, we get
\begin{align}
\Omega 
&= -  \f{S ^{\Phi} _{\>\, T}}{S^{\Phi}_{\> \Phi} +\sigma_0} +{\cal O}(\epsilon^3). 
\label{defOmega}
\end{align}
Then the angular momentum is given by
\be
J = R^2\f {\sqrt{f^+} R  - (R -3M)   \sqrt{f^-}} { \sqrt{f ^-}+ 2   \sqrt{f ^+} } \, \Omega_k \, .
\ee
Detailed calculations can be found in Ref.~\cite{uchi2}. The moment of inertia of the rotating gravastar, $I$, is defined by 
$I \equiv \epsilon J/\Omega$. Thus, we have
\be
I = {J\over\Omega_k} = \f {\sqrt{f^+} R  - (R -3M)   \sqrt{f^-}} { \sqrt{f ^-}+ 2   \sqrt{f ^+} } \, R^2 \,. \label{inertia}
\ee
Note that this definition of $I$ is different from that used in Ref.~\cite{pani2}, where $\omega$ at the surface of the gravastar is used\footnote{In the $\sigma_0=0$ limit investigated in Ref.~\cite{pani2}, $f_-=f_+$ and Eq.~\eqref{inertia} simply yields $I=MR^2$, in contrast with the value of Ref.~\cite{pani2}, $I=R^3/2$. While both definitions agree in the black-hole limit, the one adopted here has the correct dimensions of a moment of inertia. This also implies that a quantitative comparison between our results and those presented in Ref.~\cite{pani2} is not possible.} instead of $\Omega$ (cf. Eq.~\eqref{match1} in the $\sigma_0\to0$ limit). 

Note that, in the case of a tidally deformed gravastar, ${\cal O}(\epsilon)$ perturbations vanish 
because the background is nonspinning (i.e., $\omega=0$). Thus, in such case we have $J=0$ and $\Omega=0$. 

\subsubsection{Master equations for the $\epsilon^2$-order solutions of quadrupole perturbations } 
The matter three velocity of the thin shell, $u^a$, satisfies the normalization condition 
\be
h_{ab} u^a u^b = -1 \,,
\ee 
and therefore the non-zero components of $u^a$ up to $\epsilon^2$-order read 
\begin{align}
u^T &= {1\over\sqrt{f^+}} + \epsilon ^2 \left(\f {-(f^+)^{\prime} \xi^+(\Theta)+ R^2(\Omega_k - \o^+)^2\sin ^2\Theta} {2 \sqrt{f^+}^3} \right. \nn\\
& \quad\quad\quad\quad\quad\quad\quad\quad \left .-\f {h^+(R, \Theta)} {\sqrt{f^+}}\right ) + {\cal O}(\epsilon ^4)\,, \\
u^\Phi &={1\over\sqrt{f^+}}\Omega  + {\cal O}(\epsilon ^3) \,. \nn
\end{align}
The energy density of the shell, $\sigma$, is calculated through Eq. \eqref{eigen}. The explicit forms of the second order quantities $\delta \sigma_0$ and 
$\delta \sigma_2$ are given in Appendix B. The non-zero components of the projected stress tensor read
\begin{align}
\gamma^T _{\> T} &= p_0 \, q^T _{\> T}+{\cal O}(\epsilon^4), \quad \gamma^T _{\> \Phi} =p_0\,  q^T _{\> \Phi}+{\cal O}(\epsilon ^3), \nn \\  
\gamma^{\Phi} _{\> T} &=p_0 \, q^{\Phi} _{\> T}+{\cal O}(\epsilon^3) , \nn \\
\gamma ^{\Theta}_{\Theta} &=\gamma ^{+}+\gamma ^{-} \,, \quad  \gamma ^{\Phi}_{\Phi} = \gamma ^{+}-\gamma ^{-} \,, 
\label{trigam}
\end{align}
where we have defined
\begin{align}
\gamma^+ & = p_0  -\f {\epsilon ^2 } {8 \pi R^2} \left \{   \f {2(\Omega_k R^3-2 J)} {3  R^3 f^+} \left (\f{JL^2 -R^5 \Omega_k}{L^2 \sqrt{f^-}} \right.  \right .\nn \\
&  \left.+\f{ J(3 M -2 R)+M R^3 \Omega_k}{R \sqrt{f^+}} \right )-R^2 \sqrt {f^+} (h_0 ^+)^{\prime} -\f {\xi^-_0} {\sqrt{f^-} ^3}  \nn \\
& \left.+ \f {(3 M^2 -3 M R + R^2) \xi^+_0 +R(R-M) m_0 ^+} {R^2 \sqrt {f^+}^3} \right \}\nn \\
&+ \f {\epsilon ^2 P_2} {8 \pi R^2} \left \{ \f {2(J L^2 -R^5 \Omega_k)(R^3 \Omega_k -2 J^2) } {3L^2 R^3\sqrt{f^-} f^+}\right. \nn \\
&   + R^2[[ \sqrt {f} ( h_2  ^{\prime} + k_2 ^{\prime})]] \nn \\
&+\f {(3 R^2 - 2 L^2) \xi^-_2 +(L^2 - 2 R^2 ) m_2 ^-} {L^2 \sqrt {f^-} ^3} \nn \\
& +\f{2(2 J -R^3 \Omega_k) (J(3M-2 R) +M R^3 \Omega_k) }{3 R^4 \sqrt{f^+}}  \nn \\
& \left . - \f {(3 M^2 + 3 M R -2 R^2) \xi^+_2 + R(R-M) m_2 ^+} {R^2\sqrt {f^+}^3} \right \}  + {\cal O}(\epsilon ^4 ),
\end{align}
\begin{align}
 \gamma^- & = \f {\epsilon ^2 \sin ^2 \theta} {16 \pi R^2} \left \{ \f {2 (2 J - R^3 \Omega_k)} { R^3 f^+} \left (   -\f {JL^2 -R^5 \Omega_k} {L^2 \sqrt{f^-} } \right. \right. \nn \\
 & \left .\left. +\f {J(3 M -2 R) + M R^3 \Omega_k} {R\sqrt{f^+}}\right ) +3 \left[ \left[ \f {\xi_2 } {\sqrt {f}} \right ] \right] \right \}   + {\cal O}(\epsilon ^4 ). 
 \label{gammai}
\end{align}
Since we assume that the shell is composed of a perfect fluid, the projected stress tensor must be proportional to the projection tensor, i.e., $\gamma _{ab}= p q_{ab} $.
The $(\Phi,\Phi)$ and $(\Theta,\Theta)$ components of the projected tensor are 
\begin{align}
\gamma^{\Phi}_{\Phi} & = p q^{\Phi}_{\Phi}\nn \\
& =(p_0 + \epsilon^2 \delta p) (1+ u^{\Phi}u_{\Phi}) + {\cal O}(\epsilon^4) \nn \\
& =p_0 + \epsilon^2 \delta p+p_0 u^{\Phi}u_{\Phi} + {\cal O}(\epsilon^4),\\
\gamma^{\Theta}_{\Theta} & = p q^{\Theta}_{\Theta}=p_0 + \epsilon^2 \delta p+ {\cal O}(\epsilon^4).
\end{align}
Then, the condition for a perfect-fluid thin shell reduces to
\be
2\gamma^-   = -p_0 u^{\Phi}u_{\Phi}.
\ee
The explicit form of this condition is given by 
\begin{align}
\left [ \left[\f{\xi_2}{ \sqrt{f}} \right] \right ]& =\f{2}{3 \sqrt{f^+}} \left (\f{4 J^2}{R^3} - \f{2 J  (M+R) \Omega_k}{R} +  M R^2 \Omega_k ^2 \right ) \nn \\
&-2(2 J -R^3 \Omega_k) \left ( \f {4\pi R p_0} {3}  \Omega_k -\f{ J L^2 - R^5 \Omega_k }{3 L^2 R^3\sqrt{f^-}}   \right ) .
\label{xi2}
\end{align}
The explicit forms of the second order quantities $\delta p_0$ and $\delta p_2$ are given in Appendix B.
The $\epsilon^2$-order solutions for quadrupole perturbations of a slowly rotating thin-shell gravastar can be fully computed by determining the four unknown 
constants, $B$, $C_3$, $\xi_2^+$, and $\xi_2^-$, which are solutions of the set of four coupled linear algebraic equations, Eqs.~\eqref{match2}, 
\eqref{match3}, \eqref{xi2}, and $\displaystyle \delta p_2=\left({dp\over d\sigma}\right)\delta\sigma_2$ (the explicit form of $\delta p_2$ and $\delta \sigma_2$ is given in Appendix B). Note that, together with the quadrupolar perturbations, the junction conditions allow to fully determine also the spin-induced spherically symmetric deformations to ${\cal O}(\epsilon^2)$. We neglect these deformations here, details are given in Ref.~\cite{uchi2}

Let us now consider a spherically symmetric, tidally deformed gravastar. In this case, we have to impose the conditions $J=0$ and $\Omega=0$ 
in Eq.~\eqref{trigam}. Thus, the condition for the perfect fluid thin shell, Eq.~\eqref{xi2}, simply reduces to 
\be
\left [ \left [ \xi_2/\sqrt{f} \right ] \right ]=0 \,. \label{xi2t}
\ee
Note that $\epsilon^2$ in this case is just a bookkeeping parameter and can be factored out. To have a unique solution, 
we set a value of $D_2$ [see, Eq.~\eqref{love0}] as 
\be
D_2={M^2\over 3 R^2} \,. 
\label{def_D2}
\ee
The solutions with $D_2 >0$ represent the tidally deformed stars by the binary companion on the symmetry axis\footnote{We recall that, due to the spherical symmetry of the background configuration, nonaxisymmetric perturbations can be easily obtained from the axially symmetric ones, the only difference is in the angular decomposition, cf. e.g.~\cite{hinderer}. Note also that the Love numbers do not depend on the value of $D_2$ chosen in Eq.~\eqref{def_D2}, but only on the ratio $D_1/D_2$ (cf. Eq.~\eqref{love}). On the other hand, the ellipticity discussed later is proportional to the external tidal field.}
As in the spin-induced case, the $\epsilon^2$-order solutions of quadrupole perturbations for tidally deformed thin-shell gravastars can be found by determining the four unknown 
constants, $D_1$, $C_3$, $\xi_2^+$, and $\xi_2^-$, which are solutions of the set of coupled four linear algebraic equations, Eqs.~\eqref{match2}, 
\eqref{match3}, \eqref{xi2t}, and $\displaystyle \delta p_2=\left({dp\over d\sigma}\right)\delta\sigma_2$. Tidally-induced monopole terms are not given here but can be computed through the same procedure.

\section{Numerical results}
When presenting our numerical results, we often employ dimensionless quantities in terms of the length scale of the de Sitter horizon radius, $L$. In other words, we use units such that $L=1$. 
\subsection{Equation of state for the thin-shell matter}
To obtain slowly rotating and tidally deformed solutions for gravastars with a perfect-fluid thin shell, we need to assume some equation of  
state for the thin-shell matter. In Ref.~\cite{uchi2}, sequences of equilibrium solutions characterized by a constant fixed value 
of $M/L$ are considered and, from these sequences of equilibrium solutions, the corresponding equation of state is determined. In practice, for some fixed value of $M/L$, the energy density and pressure of the thin shell in an equilibrium state are, respectively,  
given by $\sigma_0=\sigma_0(R)$ and $p_0=p_0(R)$, with $R$ being the radius of the thin shell. Thus, for each value of $M/L$, we obtain a relation 
between $\sigma_0$ and $p_0$, which is used as the equation of state. The latter, however, changes with the compactness of the object.  In Ref.~\cite{pani2}, a thin shell with vanishing energy density is assumed, i.e., $\sigma_0=0$ is employed as the equation of state for the thin shell. 

In this study, we consider a more generic and realistic configuration and assume that the thin shell is composed of a polytropic fluid, whose equation of state is given by
\be
p = k \, \sigma ^{1+{1\over n}}, \label{pol_eos}
\ee
where $k$ and $n$ are positive constants and are kept fixed along a sequence of equilibrium solutions. The sound speed of the fluid, $v_s$, is defined by  
\be
v_s^2\equiv \f{dp}{d\sigma} =\left ( 1+\f{1}{n}\right ) \f{p_0}{\sigma_0} + {\cal O}(\epsilon^2).
\label{vs}  
\ee
The equation of state for the perturbation is also given by the adiabatic relation 
\be
\delta p=v_s^2\,\delta\sigma=\left ( 1+\f{1}{n}\right ) \f{p_0}{\sigma_0}\,\delta\sigma+ {\cal O}(\epsilon^2). 
\ee
Note that the speed of sound $v_s^2$ is necessary to close the system of junction conditions through $\delta p_2=v_s^2 \delta\sigma_2$, as discussed in the previous section.

\subsection{Spherically symmetric unperturbed solutions and their stability}
In this study, we construct a one-parameter family of equilibrium unperturbed gravastars for fixed equation of state, the parameter being the compactness 
of the gravastar, $c \equiv M/R$. In the unperturbed configuration, the energy density and pressure of the thin shell can be written in terms of the three physical quantities $c$, $M$, and $L$, as
\beq
\sigma_0&=  & \f{\sqrt{1-\f{M^2}{c^2L^2}}-\sqrt{1-2c}}{4 \pi M} c, \nn \\
p_0 &= & \f{c}{8 \pi M} \left ( \f{1-c}{\sqrt{1-2c}} -\f {1-\f{2M^2}{c^2L^2}}{\sqrt{1-\f{M^2}{c^2L^2}}}\right ). \label{def_p0b}
\eeq
By imposing the equation of state~\eqref{pol_eos} and by using the two above relations, we obtain an algebraic relation between $M$ and $c$. Thus, for fixed values of $n$, $k$, and $c$, 
we can compute the mass of the gravastar, $M=M(n,k;c)$. The equilibrium solution 
of the unperturbed spherical gravastar is therefore characterized by $n$, $k$, $c$, and $M$. Giving a set of fixed parameters, $n$ and $k$, we obtain sequences of equilibrium 
solutions by changing values of $c$. The effects of (spin- or tidal-)induced 
deformations of the thin-shell gravastar along several sequences of equilibrium solutions characterized by values of $k$ and $n$ are discussed later. 

Let us consider the radial stability of spherical unperturbed gravastars with a polytropic thin shell. Here, we basically rely on the dynamical stability analysis 
given by Visser and Wiltshire~\cite{vis}. In their treatment, the motion of the spherical thin shell is determined by the energy equation
\be
{1\over 2}\,\dot{R}^2+V(R)=0\,, 
\ee
where the shell radius $R$ is extended to be a function of the proper time of the thin shell, $\tau$, $\dot{R}$ denotes the proper time derivative of $R$, and 
$V$ stands for the potential, given in Eq.~(40) of Ref.~\cite{vis}. In terms of $V$ and $V' \equiv dV/dR$, the energy density and pressure of the dynamical thin shell are given by 
\begin{align}
\sigma_0&=   \f{\sqrt{1-2V(R)-\f{R^2}{L^2}}-\sqrt{1-2V(R)-\f{2M}{R}}}{4 \pi R}, \label{def_sig0d} \\
p_0 &=  \f{1}{8 \pi R} \left ( \f{1-2V(R)-RV^{\prime}(R)-\f{M}{R}}{\sqrt{1-2V(R)-\f{2M}{R}}} \right . \nn \\
& \quad\quad\quad\quad\quad \left.-\f {1-2V(R)-RV^{\prime}(R)-\f{2R^2}{L^2}}{\sqrt{1-2V(R)-\f{R^2}{L^2}}}\right ), \label{def_p0d}
\end{align}
respectively. As in a standard dynamical analysis of mechanical systems, equilibrium states satisfy $V(R)=0$ and $V^{\prime}(R)=0$ and their 
dynamical stability is determined by the sign of $V'' \equiv d^2V(R)/dR^2$ for the corresponding equilibrium solutions, namely they are stable when $V'' >0$ and unstable when $V''<0$. 
Note that Eqs.~\eqref{def_sig0d} and \eqref{def_p0d} reduce to Eqs.~\eqref{def_sig0} and \eqref{def_p0} if $V(R)=0$ and $V^{\prime}(R)=0$. 

For a polytropic thin shell satisfying Eq.~\eqref{pol_eos}, we have 
\be
\f{dp_0}{dR} = \left ( 1+\f{1}{n}\right )\f{p_0}{\sigma_0} \f{d\sigma_0}{dR}\,,
\label{deriv}
\ee
and  the above relation can be used to extract properties of the potential, $V(R)$, from the equation of sate. 

For equilibrium solutions, using Eq.~ \eqref{deriv}, we obtain  
\be
V''  = \f{W}{H}  \, ,  \label{d2v}
\ee
where
\begin{align}
W &\equiv \f{c^2}{M^2}\left \{ \f{M^2}{c^2L^2f^-}-\f{(3c-2)(1-c)}{f^+}  -\f{\sqrt{f^+}}{ \sqrt{f^-}^3} \right . \nn \\
& - \f{\sqrt{f^-} (3c^2-3c+1)}{\sqrt{f^+}^3}\nn \\
& +\left ( 1+\f{1}{n}\right ) \left [ \f{(3c-1)(1-c)}{f^+} -\f{c^2L^2-2M^2}{c^2L^2f^-}\right . \nn \\
& \left . \left.  \quad \quad \quad + \f{2\sqrt{f^+}}{ \sqrt{f^-}} - \f{2M^2(1-3c)}{ c^2L^2  \sqrt{f^+}\sqrt{f^-}} \right ] \right \} , \\
H&\equiv \f{\sqrt{f^+}}{\sqrt{f^-}}+ \f{\sqrt{f^-}}{\sqrt{f^+}}-2 .
\end{align} 
To evaluate the stability, it is sufficient to check the sign $W$, 
since $H>0$ because of the inequality of arithmetic and geometric means.  
Thus, spherical gravastars with a polytropic thin shell are stable (unstable) against radial 
perturbations when $W>0$ ($W<0$).

\begin{figure}[t] 
\includegraphics {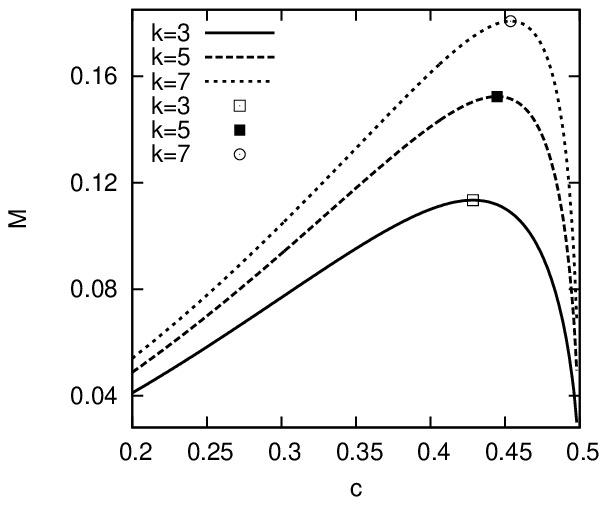}
\includegraphics {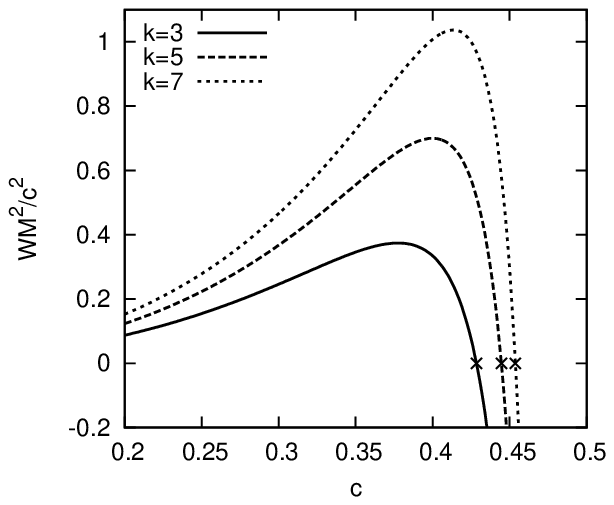}
\caption{Top: The mass $M$ of the spherical unperturbed gravastar with the $n=1$ polytropic thin shell as a function of the compactness $c$. 
Bottom: The radial stability discriminant, $W M^2/c^2$, for the spherical unperturbed gravastar with the $n=1$ polytropic thin shell
as a function of the compactness $c$. Each curve corresponds to the sequence of spherical unperturbed gravastars characterized
by the same values $k$. The marginally stable solutions are indicated by marks (square, circle, or cross mark).}
\label{equili}
\end{figure}

\begin{figure}[t] 
\includegraphics {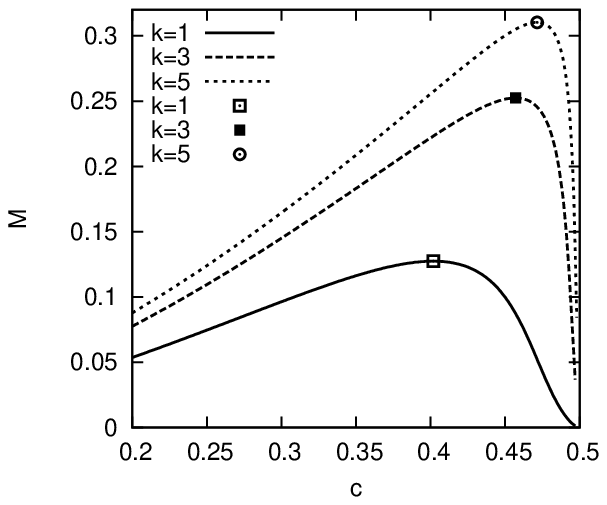}
\includegraphics {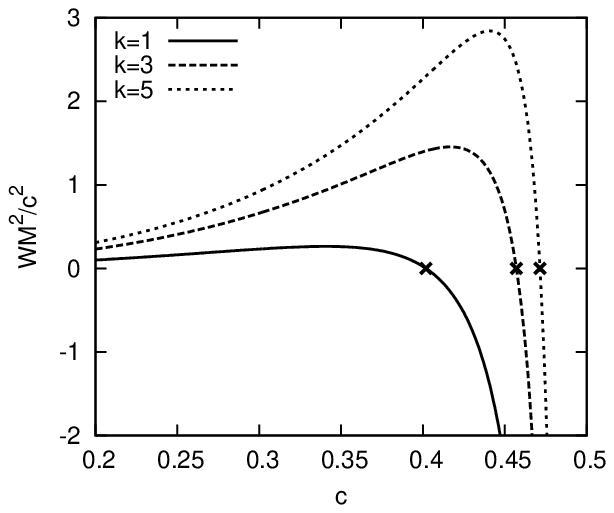}
\caption{Same as Fig.~\ref{equili}, but for the spherical unperturbed gravastar with the $n=3$ polytropic thin shell.}
\label{equili-n3}
\end{figure}

Figures~\ref{equili} and \ref{equili-n3} show three equilibrium sequences of unperturbed spherical gravastars characterized 
by $(n,k)=(1,3)$, $(1,5)$, and $(1,7)$ and by $(n,k)=(3,1)$, $(3,3)$, and $(3,5)$, respectively. Later on, we present the results for the spin-induced and tidal deformations for these six choices of equations of state. In these figures, 
the behaviors of $M$ and $W M^2/c^2$ are shown as functions of $c$. 
As shown in the top panels of Figs.~\ref{equili} and \ref{equili-n3}, the mass displays a local maximum as a function of the compactness along the sequences of equilibrium solutions. 
The maximum masses are, respectively, given by $M \approx 0.11344$, $0.15239$, and $0.18070$ for the sequences with $(n,k)=(1,3)$, $(1,5)$, 
and $(1,7)$, and by  $M \approx 0.12744$, $0.25245$, and $0.31035$ for the sequences with $(n,k)=(3,1)$, $(3,3)$, and $(3,5)$. 
The marginally stable solutions characterized by $W=0$ appear at $c \approx 0.42843$, $0.44461$ and $0.45372$ for the sequences 
of solutions with $(n,k)=(1,3)$, $(1,5)$, and $(1,7)$, and at $c \approx 0.40183$, $0.45705$ and $0.47146$ for the sequences 
of solutions with $(n,k)=(3,1)$, $(3,3)$, and $(3,5)$ (see the bottom panels of Figs.~\ref{equili} and \ref{equili-n3}). 
In Figs.~\ref{equili} and \ref{equili-n3}, the marginally stable solutions are indicated by marks (square, circle, or cross mark, depending on the equation of state). Here and henceforth, 
the square, circle, and cross marks are used to indicate the marginally stable solutions. As shown in the top panels of Figs.~\ref{equili} and 
\ref{equili-n3}, these marginally stable solutions coincide with the maximum mass solutions for the equilibrium sequences characterized by fixed 
values of $n$ and $k$ within the accuracy of the numerical calculation. 
Thus, the critical point of the radial stability for the spherical thin-shell gravastar seems to coincide with the maximum mass solutions along the sequences of equilibrium 
solutions characterized by a single equation of state, or fixed values of $n$ and $k$, which is the usual situation for generic relativistic polytropic spheres. 
Although this result strongly suggests that the same stability criterion that applies to ordinary self-gravitating polytropic fluids also works for thin-shell gravastars, it would be interesting to find a generic proof for this behavior.

Finally, let us discuss the energy conditions for the thin shell.
We can get the upper limit of gravastar mass $M_c$ to satisfy the dominant energy condition by squaring $\sigma_0 - |p_0| \geq 0$ twice.
The square of the upper limit $M_c^2$ is given as 
\be
\begin{split} 
\f{M_{c}^2 }{L^2}&= c^2 \left (\f{3}{4} -\f{{\cal G}^2}{8} -\f{\sqrt{{\cal G}^4+4{\cal G}^2}}{8}\right ),\\
{\cal G} &= \sqrt{1-2c} +\f{1-c}{2\sqrt{1-2c}}.
\end{split}
\ee
The explicit critical values of the compactness for $\sigma_0-|p_0| =0$ can be obtained from the intersection points of $M_c$ and $M(c) $ in the top panels in Figs. \ref{equili} and \ref{equili-n3}, and they are $c \approx 0.47813$, $0.47459$ and $0.46877$ for the sequences 
of solutions with $(n,k)=(1,3)$, $(1,5)$, and $(1,7)$, respectively, and $c \approx 0.47979$ and $0.27241$ for the sequences 
of solutions with $(n,k)=(3,1)$ and $(3,3)$, respectively.
The solution with $(n,k)=(3,5)$ does not satisfy the dominant energy condition in the range of $ 0.2 \le c \le 0.5$. 
Thus, except for the solutions with $(n,k)=(3,3)$ and $(n,k)=(3,5)$, radially stable solutions satisfy the dominant energy condition.

\subsection{Rescaled physical quantities expressing rotational and tidal deformations} 
Yagi and Yunes~\cite{yagi,yagi2} have found that for neutron star models and quark star models, the dimensionless quantities 
\be
\bar{I} \equiv \f{I}{M^3}\,, \quad \bar{\lambda}\equiv \f{\lambda}{M^5} \,, \quad \bar{Q} \equiv \f {QM}{J^2}  \,, 
\ee
obey universal relations within an accuracy of a few percent for a variety of reasonable equations of state. 

As mentioned before, here we are interested in whether similar nearly universal relations among $\bar{I}$, $\bar{\lambda}$, and $\bar{Q}$ exist also for thin-shell gravastars, and we wish to investigate their behaviors in the black-hole limit. 
For slowly rotating Kerr black 
holes, the angular momentum $J$ and angular velocity at horizon $\Omega_H$ are, respectively, given by $J=a\,M_{\rm BH}$ and 
$\displaystyle \Omega_H={a\over r_+^2 +a^2} = {a\over 4 M_{\rm BH}^2}+{\cal O}(a^3)$, where $a$, $M_{\rm BH}$, and $r_+$ are the spin parameter, mass, and 
event-horizon radius of the black hole, respectively. Therefore, the moment of inertia reads 
$\displaystyle I_{\rm BH} ={J\over\Omega_H}= 4 M_{\rm BH}^3+{\cal O}(a^2)$. Thus, for slowly rotating Kerr black holes, we have $\bar{I} =4 +{\cal O}(a^2)$. Furthermore, as argued 
by Binnington and Poisson~\cite{bin}, the tidal Love numbers of a nonrotating black hole are zero. Thus, we have $ \bar{\lambda}=0$. Finally, from 
Eq.~\eqref{def_Q}, the dimensionless quadrupole moment, $\bar{Q}$, is given by
\be
\bar{Q}  = 1 + \f{8BM^3}{5J^2}.
\ee
As mentioned before, regularity at the event horizon of a slowly rotating Kerr black hole implies $B=0$ and, therefore, $\bar{Q}=1$ in the black-hole limit.  

On the other hand, for standard 
neutron-star models, the centrifugal force makes the star oblate, which corresponds to $B>0$. Thus, $\bar{Q}>1$ for 
standard rotating neutron stars. This is not the case for the slowly rotating thin-shell gravastars as discussed in Refs.~\cite{uchi2,pani2} and below. 
\subsection{Rotational and tidal deformations of thin-shell gravastars}
Let us first examine the deformation of the thin shell on the surface of the gravastar. Such deformation of the spherical thin shell 
due to centrifugal and tidal forces may be well described by the square of the ellipticity, defined by   
\be
\f{e^2}{\epsilon^2} \equiv  -3 \left ( k_2+\f{\xi_2}{R}\right ) +{\cal O}(\epsilon^2) \,. 
\label{eq-e2}
\ee
Here, we use the standard definition of the ellipticity for rotating stars given in, e.g., Ref.~\cite{thorne}. Thus, we have $e^2 >0$ ($e^2 <0$) for 
oblate (prolate) spheroids.   
\begin{figure}[t] 
\includegraphics {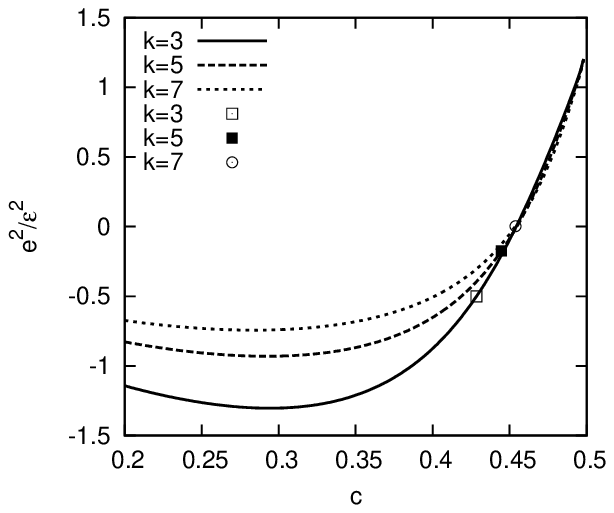}
\includegraphics {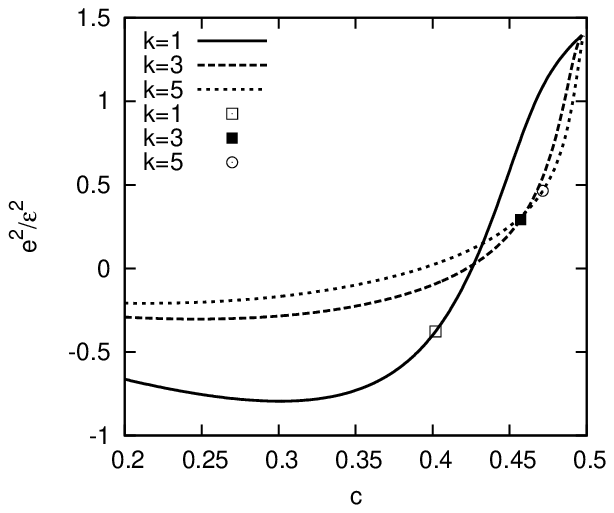}
\caption{The square of ellipticity, $e^2/\epsilon^2$, of the thin shell as function of the compactness $c$ for slowly rotating gravastars with $n=1$ (top panel) and $n=3$ (bottom panel) polytropic 
thin shells. Each curve corresponds to the sequence of spherical unperturbed gravastars 
characterized by the same values of $k$. The marginally stable solutions are indicated by marks (square and circle).}
\label{e2}
\end{figure}
\begin{figure}[t] 
\includegraphics {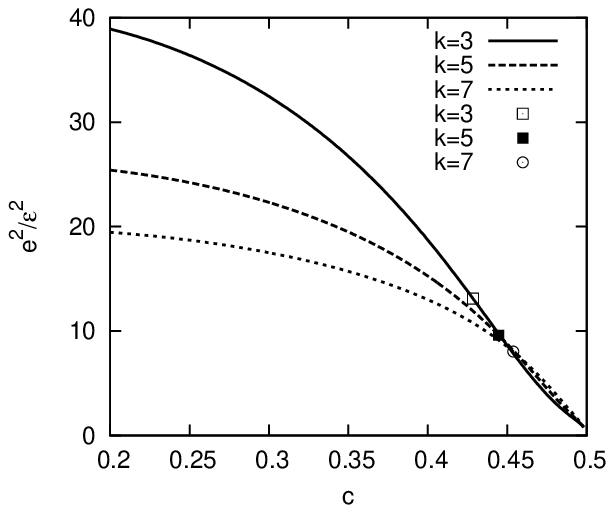}
\includegraphics {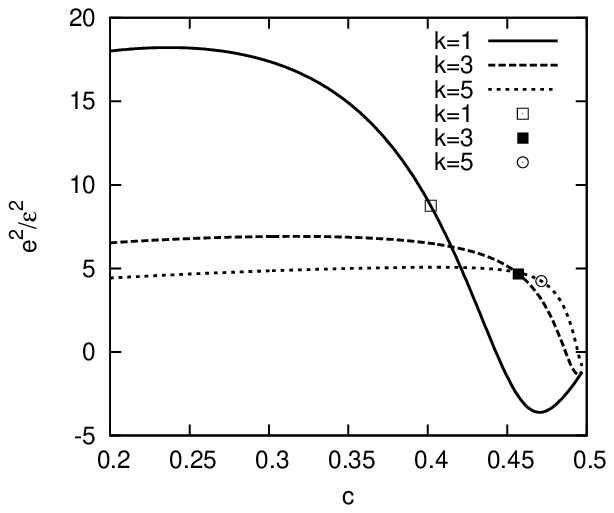}
\caption{Same as Fig.~\ref{e2}, but for the tidally deformed gravastars.}
\label{et}
\end{figure}

In Figs.~\ref{e2} and \ref{et}, the squares of the ellipticity of the thin shell respectively	associated with the rotational and tidal deformations for 
the six sequences of unperturbed spherical gravastar solutions previously presented are 
given as functions of the compactness $c$. In each figure, the results for the $n=1$ and $n=3$ polytropic thin shells are presented 
in the top and bottom panels, respectively.  

For the rotationally deformed thin shell, the square of the ellipticity is negative for less compact models, changes sign 
as the compactness increases, and finally converges to a single positive value in the nearly black-hole limit (see Fig.~\ref{e2}).
As shown in Fig.~\ref{e2}, 
the values and the behavior of $e^2/\epsilon^2$ in the nearly black-hole limit strongly depend on the polytrope index $n$. In particular, the ellipticity is generically discontinuous in the black-hole limit, as observed in Ref.~\cite{pani2} for the $\sigma_0=0$ case. 
For ordinary rotating perfect-fluid stars,  the ellipticity at the stellar surface is always positive, which reflects the fact that the centrifugal force typically makes the star 
oblate. As discussed before and in Refs.~\cite{uchi2,pani2}, the situation for the thin-shell gravastar is different from that for the fluid star, i.e., the centrifugal force makes the thin 
shell gravastar prolate at small compactness. Although this is counterintuitive, a similar situation is also observed in the case of rotating stars with anisotropic pressure~\cite{yagi5}. As 
argued in Ref.~\cite{yagi5}, we think that a key element is the strongly anisotropic stress of the thin-shell matter (for the case of the thin shell, strongly anisotropic stress 
means the absence of the radial pressure) though the reason for occurrence of rotating prolate gravastars is not yet clear. A different explanation is given in Ref.~\cite{pani2} in terms of the peculiar $p=-\rho$ equation of state of the de Sitter interior\footnote{We thank Leo Stein for suggesting this explanation.}.

The case of tidal deformations is slightly more complicated than that of spin-induced deformations (cf. Fig.~\ref{et}). For a polytropic thin shell with $n=1$, (cf. top panel of Fig.~\ref{et}), the ellipticity is a positive, decreasing monotonic function of the compactness, and converges to a single 
value in the nearly black-hole limit.  For a polytropic thin shell with $n=3$ (cf. bottom panel of Fig.~\ref{et}), the ellipticity is positive 
for less compact models, changes sign as the compactness increases, and finally converges to a single negative value in the nearly black-hole limit. For 
the $n=3$ case, behavior of $e^2/\epsilon^2$ is very different from, and more complicated than, the $n=1$ case. 

For tidally deformed perfect-fluid stars, the ellipticity at the 
stellar surface is always negative, which means that the tidal force in the present setup makes a perfect fluid star prolate (see the arguments given at the end 
of Sec.~II~C). On the other hand, for tidally deformed thin-shell gravastars, the direction of the elongation of the thin shell on the meridional cross section is turned by $90$~degrees relative to the standard case. The reason for occurrence of this unusual effect is likely the same as that for the case of the rotationally deformed prolate 
thin shell, i.e., two key elements are the strongly anisotropic stress of the thin-shell matter (cf. Ref.~\cite{yagi5}) and the peculiar equation of state of the de Sitter fluid making the gravastar interior~\cite{pani2}.

\subsection{I-Love-Q relations for gravastars with polytropic thin shell}

\begin{figure}[t] 
\includegraphics {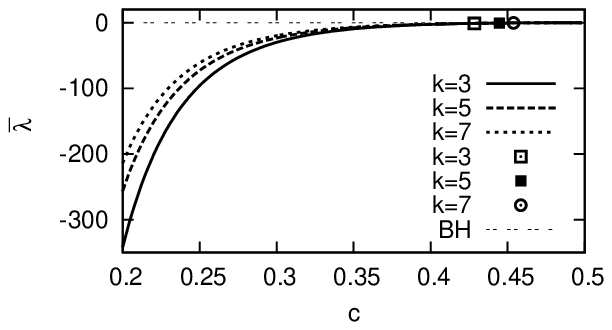}
\includegraphics {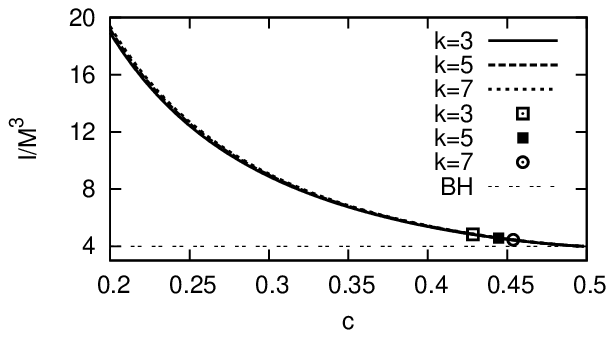}
\includegraphics {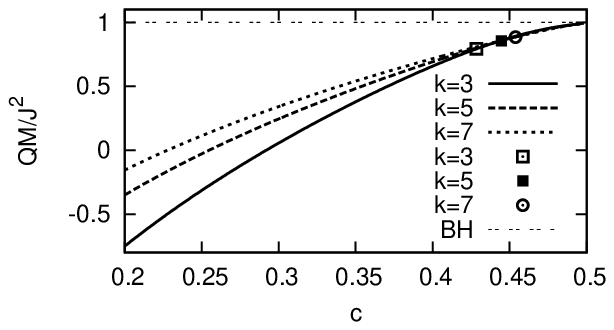}
\caption{The dimensionless tidal Love number, $\bar{\lambda}$ (top panel), the dimensionless moment of inertia, $\bar{I}$ (middle panel), and the dimensionless quadrupole moment, $\bar{Q}$ (bottom panel), as functions of the compactness $c$ for gravastars with $n=1$ polytropic thin shell. Each curve corresponds to the sequence of spherical unperturbed gravastars characterized by the same values of $k$. 
The marginally stable solutions are indicated by marks (square and circle).}
\label{n1}
\end{figure}
\begin{figure}[t] 
\includegraphics {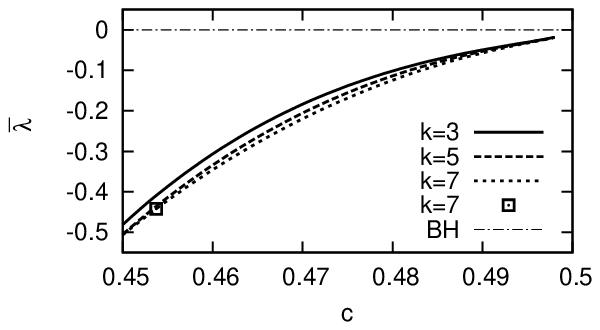}
\includegraphics {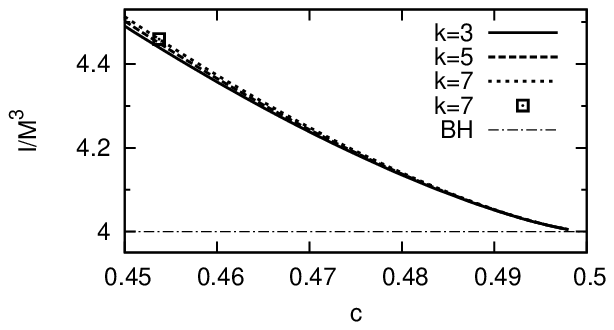}
\includegraphics {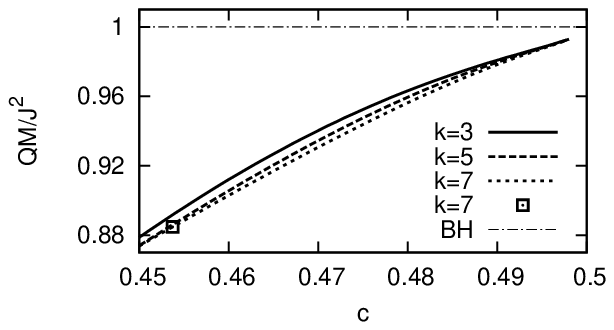}
\caption{Same as Fig. \ref{n1}, but for the strongly relativistic regime ($0.45 \le c \le 0.5$).}
\label{n1-mag}
\end{figure}
\begin{figure}[h] 
\includegraphics {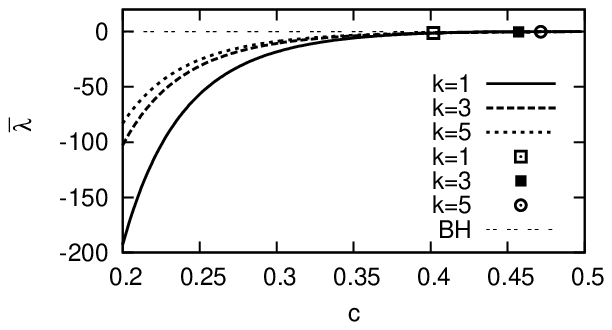}
\includegraphics {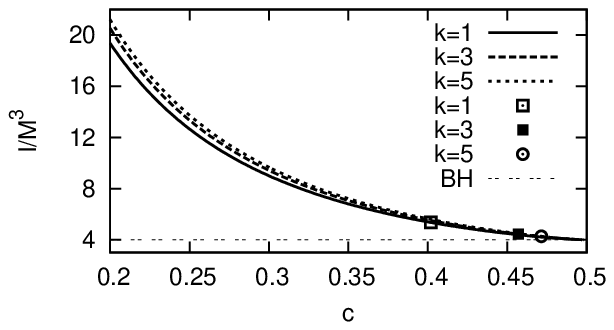}
\includegraphics {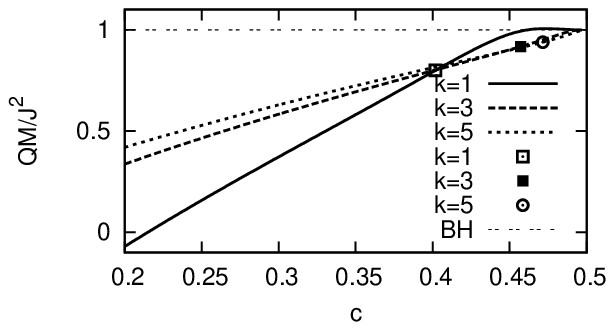}
\caption{Same as Fig.~\ref{n1}, but for the gravastar with the $n=3$ polytropic thin shell. 
}
\label{n3}
\end{figure}
\begin{figure}[t] 
\includegraphics {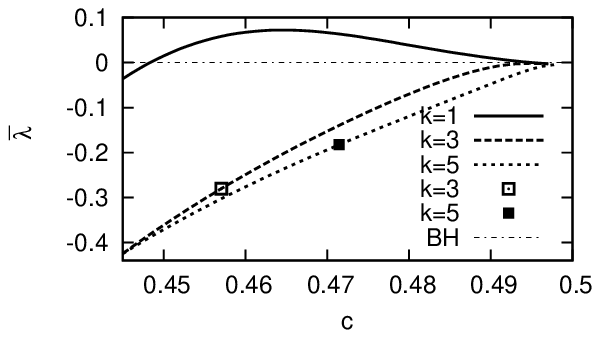}
\includegraphics {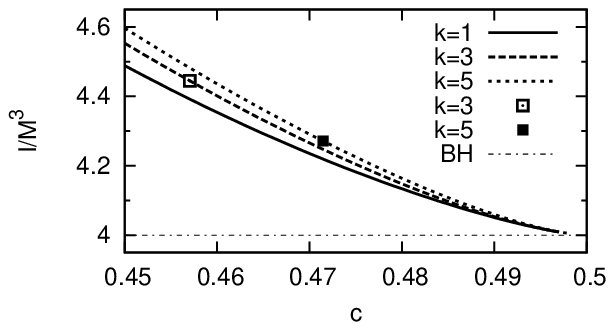}
\includegraphics {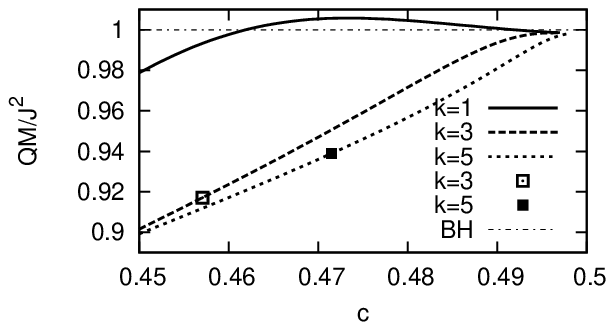}
\caption{Same as Fig.~\ref{n3}, but for the strongly relativistic regime ($0.44 \lesssim c \le 0.5$).}
\label{n3-mag}
\end{figure}
Let us now examine how the I-Love-Q triad depends on the compactness. The quantities $\bar{\lambda}$, $\bar{I}$, and $\bar{Q}$ are shown as functions of $c$ in Figs.~\ref{n1} and \ref{n1-mag} and Figs.~\ref{n3} and \ref{n3-mag} for the sequences of the 
spherical gravastars with the $n=1$ and $n=3$ polytropic thin shells, given in Figs.~\ref{equili} and \ref{equili-n3}, respectively. Figures.~\ref{n1-mag} 
and \ref{n3-mag} are zoomed-in versions of the same quantities of Figs.~\ref{n1} and \ref{n3} in the range $0.45 \lesssim c \le 0.5$, in order to better visualize their behavior in the nearly black-hole limit. In each figure, we show the corresponding quantities for the case of a slowly rotating Kerr black hole with dash-dotted or dotted horizontal lines.  

From the middle panels of Figs.~\ref{n1}--\ref{n3-mag}, we see that the dimensionless moment of inertia $\bar{I}$ is always positive, and monotonically decreases to the black-hole value of $\bar{I}$ as the compactness increases. These basic properties seem to be common also for other slowly rotating objects, e.g. slowly rotating neutron stars. We also 
observe that the function $\bar{I}(c)$ is only mildly dependent on the equation of state for the thin-shell matter. 

The functions $\bar{Q}(c)$ and $\bar{\lambda}(c)$ are respectively given in the bottom and top panels of Figs.~\ref{n1}--\ref{n3-mag}. For $n=1$, we find $\bar{Q}<1$ and $\bar{\lambda}<0$, which is related to the gravastar being prolate for rotational deformations and oblate for tidal deformations. Note that a similar behavior is observed for thin-shell gravastar with 
$\sigma_0=0$~\cite{pani2}. This property is consistent with the fact that less compact models of slowly rotating (tidally deformed) thin-shell 
gravastar have a prolate (oblate) shaped thin shell, as previously discussed. From the bottom (top) panels of Figs.~\ref{n1}--\ref{n3-mag}, we also see that 
$\bar{Q}(c)$'s ($\bar{\lambda}(c)$'s) obtained in this study are basically monotonically increasing functions which approach the value corresponding 
to the black-hole limit as $c\to1/2$. 
An exception appears in the case of the very compact models with $(n,k)=(3,1)$ for $c \gtrsim 0.45$. For this exceptional 
sequence of unperturbed spherical gravastars, $\bar{Q}$ ($\bar{\lambda}$) exceeds unity (zero) around $c\sim 0.46$ 
($c\sim 0.45$), but once again becomes less than unity (zero) around $c\sim 0.49$ ($c\sim 0.495$) (cf. bottom (top) panel of Fig.~\ref{n3-mag}). 
In other words, in this case $\bar{Q}(c)$ and $\bar{\lambda}(c)$ cross the black-hole value twice, and then approach the black-hole limit.

\begin{figure}[t] 
\centerline{\includegraphics[width=0.47\textwidth]{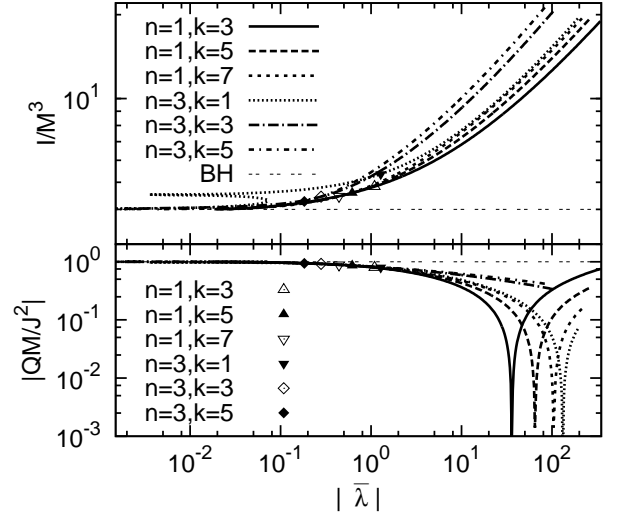}}
\caption{
 The $\bar{I}$--$|\bar{\lambda}|$ (top panel) and the $|\bar{Q}|$--$|\bar{\lambda}|$ (bottom panel) relations for a gravastar with a polytropic thin shell. The marginally stable solutions are indicated by marks (triangle and diamond).}
\label{iloveq}
\end{figure}
In Fig.~\ref{iloveq}, we present the I-Love-Q relations for gravastars with a polytropic thin shell. Top and bottom panels correspond to the $\bar{I}$--$|\bar{\lambda}|$ and the $|\bar{Q}|$--$|\bar{\lambda}|$ relations, respectively. 
We observe various interesting properties. First, the I-Love-Q relations for thin-shell gravastars are drastically different from those of neutron stars and quark stars. For example, $\bar{\lambda}$ and $\bar{Q}$ for the gravastar case can be negative (note that we use $|\bar{\lambda}|$ and $\left | \bar{Q}\right |$ instead of $\bar{\lambda}$ and $\bar{Q}$ to represent the I-Love-Q relations in Fig.~\ref{iloveq}).
Second, we confirm that the relations approach their corresponding black-hole limit~\cite{pani2} as the compactness increases (or, equivalently, as $|\bar{\lambda}|$ decreases). This property seems to hold in general, i.e. for all values of $k$ and $n$ considered in this study.

Finally, we observe that the I-Love-Q relations for thin-shell gravastars substantially  
depend on the equations of state for the thin shell. Although all curves approach the same black-hole limit, they behave differently for different polytropic indices $n$ and even for the same polytropic index but different values of $k$. Therefore, there seems to be 
no universality in the I-Love-Q relations for thin-shell gravastars. However, even in the cases of neutron stars and quark stars the universality does not hold for \emph{any} equation of state, but only for a subclass of equations of state which nonetheless comprises the majority of models, ranging from soft to stiff matter. These models differ at very high densities but are similar near the neutron-star surface, where the behavior of matter is well understood. In the case of thin-shell gravastars, the equation of state changes precisely at the surface and, in this sense, it is reasonable to expect a different behavior. 
It would be interesting to extend our study to the case of finite-thickness gravastars, in order to investigate the region which is (possibly) most responsible for the emergence of the universality.

In the present study, we allow for a wide range of polytropic equations of state for the thin shell because the properties of the latter are basically unconstrained. It is therefore natural to expect larger deviations in the universality of the I-Love-Q relations relative to the neutron star case (whose equation of state, as a qualitative reference, can be roughly approximated by a polytrope with $0.5\lesssim n \lesssim1$). In fact, we expect even larger deviations from universality for polytropic indices $n>3$. If the 
equation of state for the thin shell is moderately restricted, therefore, it is possible that the I-Love-Q relations for thin-shell gravastars will become moderately universal, the degree of the universality would depend on the class of equations of state considered. 

\section{Gravitational-wave constraints on the tidal deformability of thin-shell gravastars}
Since the tidal Love numbers of static~\cite{bin,Gurlebeck:2015xpa} (and, presumably, also rotating~\cite{pani3a,pani3}) black holes are identically zero, any measurement of a nonzero tidal Love number of a compact object implies that the latter is not a black hole. This fact is particularly important for gravitational-wave tests of black-hole mimickers and of exotic compact objects. In the near future, gravitational-wave detections of compact-binary inspirals~\cite{GW150914,GW151226} will be routine; such observations may be used to put upper bounds on the tidal Love number of the two bodies, thus constraining exotic alternatives.

The first post-Newtonian corrections to the inspiral phase of two compact bodies depend only on their masses and spins, but not on their internal structure. For example, the chirp mass ${\cal M}\approx 30\,M_\odot$ of the event GW150914 places a lower bound on the total mass of the system, $m:=m_1+m_2\gtrsim 70\,M_\odot$. On the other hand, the merger frequency indicates that the two bodies have to be very compact~\cite{GW150914}. In other words, the objects that produced the gravitational-wave events detected so far are (at least) as compact as a neutron star but much more massive. This suggests a pair of merging black holes as the most natural source of the gravitational-wave signal~\cite{GW150914}. 

Tidal deformations enter the gravitational waveform formally at fifth post-Newtonian order, although they are enhanced by terms proportional to $c^5$~\cite{flanagan,hinderer}. In the stationary phase approximation, the tidal contribution to the gravitational-wave phase reads~\cite{flanagan}
\begin{equation}
 \delta \Psi=-\frac{9}{16}\frac{v^5}{\mu m^4}\left[\left(11m_2+m\right)\frac{\lambda_1}{m_1}+ 1 \leftrightarrow 2\right]\,,
\end{equation}
where $v$ is the orbital velocity, $m$ and $\mu$ are the total and reduced mass of the binary, respectively. Clearly, this term is zero if the two objects are black holes, because $\lambda_1=\lambda_2=0$. 

Thus, an interesting question is to what extend a putative upper limit on $\lambda$ can constrain gravastar models. 
The merger frequency and individual masses of GW150914 suggest a lower limit on the compactness of the two bodies, $c\gtrsim 0.25$. Because $\lambda$ vanishes in the black-hole limit, an upper bound on $\lambda$ can be converted into a lower bound on $c$. From our results in Figs.~\ref{n1} and \ref{n3}, we observe that $|\bar\lambda|\lesssim 20-100$ when $c\gtrsim 0.25$, the exact value depending on the equation of state of the thin shell\footnote{In the large compactness limit, the Love number is well fitted by the expression $\bar\lambda(c)\sim a(c-1/2)^b$, where $a$ and $b$ are constants that depends on the equation of state.}. Therefore, an upper bound at the level of
\begin{equation}
 |\lambda_i| \lesssim (1-7)\times 10^{-18} \left(\frac{m_i}{30\,M_\odot}\right)^5 \,{\rm s}^5\,,
\end{equation}
would provide a lower bound on $c_i:=m_i/R_i$ which is more stringent than the approximate one set by the merger frequency.

A detailed analysis of the possible gravitational-wave constraints on $\lambda$ goes beyond the scope of this work. Furthermore, because upper limits on the tidal deformability of the binary systems of GW150914 and GW151226 have not been published yet, it is difficult to estimate the constraints on $\lambda_i$ for these systems. However, similar analyses for neutron-star binary systems have been performed before the recent gravitational-wave detection (cf., e.g., Ref.~\cite{DelPozzo:2013ala}). By assuming similar absolute errors, one might expect an observational bounds at the level of $|\lambda|\lesssim 10^{-23}\,{\rm s}^5$ which, for an object of $M\approx 30M_\odot$, corresponds to $|\bar\lambda|\lesssim10^{-4}$. This would provide a very stringent lower bound on the compactness and can potentially rule out several equations of state for which a very compact thin-shell gravastar is radially unstable. In fact, as clear from the results previously presented, all models of gravastars considered in this paper would be radially unstable when $|\bar\lambda|\lesssim10^{-4}$. 

\section{Conclusion}
We have studied the rotational and tidal quadrupole deformations of a thin-shell gravastar, which are characterized by the tidal Love number and the rotational quadrupole moment, respectively. We worked in a perturbative regime in which the tidal and the rotational effects are described by small deviations from spherical symmetry.
We considered a thin shell made of a polytropic fluid and studied in detail the cases when the polytropic indices are $n=1$ and $n=3$.
We found that the I-Love-Q relations of a thin-shell gravastar are drastically different from those of an ordinary compact star 
like a neutron star. The Love number and quadrupole moment are negative for less compact models and  
the I-Love-Q relations continuously approach the black-hole limit. 

The appearance of negative values of the Love number and quadrupole moment 
means that the elongation direction of the matter distribution on the meridional cross section is turned by $90$~degrees relative to the case of an ordinary compact star. The reason for the appearance of this counterintuitive deformation is not entirely clear, 
but seems to be related to the strongly anisotropic stress of the thin shell in the horizontal and vertical directions, and to the peculiar equation of state of the gravastar's interior. Similar counterintuitive properties have been found also for others infinitesimally-thin shells deformed by rotational or tidal effects~\cite{dlc,pf,pf2,pani2,uchi2}. 

We considered a wide range of polytropic equations of state for the thin-shell matter. Within this range, there is no universality in the I-Love-Q relations, unlike the case of neutron stars and quark stars for which such relations depend only mildly on the stellar equation of state, at the level of a few percent. Although (some degree of) approximate universality might be restored by restricting to a subclass of equations of state, it is nevertheless interesting that thin-shell gravastars provide an example of very compact objects for which the universality of the I-Love-Q relations is manifestly broken.

We also evaluated the stability of the unperturbed spherical gravastars with a polytropic perfect-fluid thin 
shell against small radial perturbations. It is found that less compact models are stable and that the stability changes at the maximum mass models 
of equilibrium sequences characterized by a single equation of state. The stability analysis implies that the gravastars become unstable as 
they approach the black-hole limit. Therefore, we inevitably had to use unstable models to investigate the I-Love-Q relations very close to the black-hole limit. The instability region shrinks to zero as $k$ increases for a fixed value of $n$. The $k\to\infty$ limit corresponds to the case of a thin shell with vanishing energy density studied in Ref.~\cite{pani2}. Despite some differences in our analysis (we corrected the definition of the moment of inertia used in Ref.~\cite{pani2} and kept the properties of the thin-shell matter fixed along different sequences of solutions) our results are qualitatively similar to those presented in Ref.~\cite{pani2} and extend the latter to more realistic configurations.

In this study, we focus on the case of a perfect-fluid thin shell. However, a thin shell with anisotropic pressure might be more reasonable. For example, if the phase transition replacing the would-be horizon~\cite{gravastar} is due to a scalar field, the latter would naturally give rise to anisotropic stresses. Thus, it would be interesting to investigate in detail the case of gravastars with an anisotropic thin shell. Likewise, it would be interesting to extend our computation to the case of finite-thickness shells, although that would likely require a numerical integration of the field equations within the shell. 

A detailed analysis of the observational constraints on the tidal deformability of gravastars coming from gravitational-wave measurements goes beyond the scope of this work. Nonetheless, our results suggest that near-future constraints on the tidal Love numbers of compact objects from inspiral gravitational waveforms can place very stringent lower limits on the compactness of the two objects, ruling out several gravastar models and giving further support to the fact that events like GW150914 and GW151226 are coalescences of a pair of black holes.

So far, studies of rotating models of gravastars are restricted 
to the case of slow rotation. On the other hand, it is likely that a substantial fraction of black holes are rapidly spinning, and this is surely the case for black holes formed in the coalescences recently detected in the gravitational-wave band by the LIGO interferometer~\cite{GW150914,GW151226}. Highly-spinning gravastars might be unstable against the ergoregion instability~\cite{Cardoso:2007az}, and a detailed studied is therefore necessary to assess their viability as exotic compact objects and black-hole mimickers. These investigations remain as future work.

\begin{acknowledgments}
N.~U. acknowledges financial support from MEXT Grant-in-Aid for Scientific
Research on Innovative Areas ``New Developments in Astrophysics
Through Multi-Messenger Observations of Gravitational Wave Sources''
(Grant Number 24103005). This project has received funding from FCT-Portugal through the project IF/00293/2013.
\end{acknowledgments}

\appendix
\section{The second order coefficients of the extrinsic curvature}
%
If we expand the extrinsic curvature as ${K}_{ab}={}^{(0)}{K}_{ab}+\epsilon \, {}^{(1)}{K}_{ab}+\epsilon^2 \, {}^{(2)}{K}_{ab}$, the second order 
coefficients of the extrinsic curvature associated with the thin shell, $ {}^{(2)}{K}_{ab}$ is given as follows: 
\begin{align}
^{(2)}{K}^T_T &=  \f{1}{ R \sqrt{f^+}^3 }   \Big[ 2m_0 f^{\prime} -4R f^2 h_0^{\prime} \nn \\
& +R\xi_0\left ((f^{\prime})^2-\f{4f(f-1)}{R^2} \right )+\f{4R^3}{3} \o \o^{\prime}\nn \\
 & +\Big \{2m_2 f^{\prime}-3 R f^2 h_2^{\prime}  \nn\\
&\left. \left . +R\xi_2\left ((f^{\prime})^2-\f{4f(f-1)}{R^2} \right )-\f{4R^3}{3} \o \o^{\prime} \right \}P_2  \right ] ,\\
^{(2)} {K}^{\Theta}_{\Theta} &+^{(2)}{K}^{\Phi}_{\Phi} \nn \\
& = \f {1} { 3R^2 \sqrt {f}} \Big [ 6m_0+3\xi_0(2 f -Rf^{\prime}) -R^4 \o \o^{\prime}  \nn \\ 
&- \Big \{6m_2 +3 \xi_2 (2 f -Rf^{\prime}-6 ) -6 R^2 f k_2 ^{\prime}\nn \\
& + R^4 \o \o^{\prime} \Big \} P_2 \Big ] ,\\
^{(2)} {K}^{\Theta}_{\Theta} &- ^{(2)}{K}^{\Phi}_{\Phi}  = \f {6\xi_2+R^4 \o \o^{\prime}} { 2 R^2 \sqrt {f}} \sin^2 \Theta, \\
 ^{(2)}K &= \f{1}{R^2 \sqrt{f}^3 } \Big [ 2m_0(4f+Rf^{\prime})-4R^2 f^2 h_0^{\prime}\nn \\
  & +\xi_0 \left(4f(1+f-Rf^{\prime})+R^2 (f^{\prime})^2 \right ) \nn \\
  &+ \Big \{ 2 m_2 (4f+R f^{\prime})-4 R^2 f^2 (h_2^{\prime}+2k_2^{\prime})   \nn \\
  &+\xi_2 \left (4 f (-5+f-R f^{\prime})+R^2 (f^{\prime})^2 \right) \Big \} P_2 \Big ]. 
\end{align} 
\section{The second order energy density and pressure perturbations of the thin shell}
%
The $l=2$ energy density and pressure perturbations are, respectively, given by 
\begin{align}
\delta \sigma_2 & = \f{1}{4 \pi R^2} \left \{ \f{2 \xi^-_2 -m_2^-}{\sqrt{f^-} }-\f{(2 J -R^3 \Omega_k)^2}{3 R^3 f^+\sqrt{f^-}}  - R^2 [[\sqrt{f} k_2^{\prime}]]\right. \nn \\
& +\f{J(-2 J+R^3 \Omega_k)+R^3 m_2}{R^3\sqrt{f^+}} -\f{3M+2R}{R\sqrt{f^+}}\xi_2^+ \nn \\
& \left. +\f{(R^3 \Omega_k -2 J)(J +R^2 (R-3 M)\Omega_k)}{3 R^3 \sqrt{f^-}^3 }   \right \},
\end{align}
\be
\begin{split}
\delta p_2 & = \f{1}{8\pi R^2} \left \{R^2 \left [ \left[ \sqrt{ f} (h_2^{\prime} + k_2 ^{\prime})  \right ] \right ] \right. \\
& +\f{1}{R\sqrt{f^+}^3} \left ( \f {(2 R^2 -3 M R +3 M^2) \xi^+_2 }{R}-(R-M) m_2 ^+ \right )\\
& + \f{R^3 \Omega_k-2 J}{3 R\sqrt{f^+}^3} \left( \f{ 2 J(2 R-3 M)}{R^3} + (R-3M) \Omega_k) \right)\\
& \left .-\f {(R^3 \Omega_k -2 J) ^2} {3R^3 f^+ \sqrt{f^-}}+\f {(L^2 -2 R^2) m_2^--(2 L^2 -3 R^2) \xi^-_2} {L^2 \sqrt{f^-} ^3} \right \} .
\end{split}
\ee
The $l=0$ energy density and pressure perturbations are, respectively, given by
\begin{align}
\delta \sigma_0 & = \f{1}{4 \pi R^5} \left \{  \f{( \Omega_k -2 J)^2}{3  f^+ \sqrt{f^-}}\right .+ \f{1}{ \sqrt{f^+}} \Big ( J (2 J-R^3 \Omega_k)  +  m_0^+ \nn \\
& \left . \left. +\f{(2 J -R^3 \Omega_k)(J+R^4 (R-3M)\Omega_k)}{3 f^+} \right ) \right \} \nn \\
&-\f{2(\sigma_0 +p_0)}{R}\xi_0,
\end{align}
\begin{align}
\delta p_0 & = -\f {1} {8 \pi R^2} \left \{\f{(3 M^2 -3 M R +R^2)\xi_0 +R(R-M)m_0}{R^2 \sqrt{f^+}^3}\right. \nn \\
& + \f {(R^3 \Omega_k -2 J) (2 J (2 R-3 M)+R^2 (R-3M) \Omega_k)} {3 \sqrt{f^+} R^3}  \nn \\
&\left.- \f {(2 J - R^3 \Omega_k)^2} {3 R^3 f^+ \sqrt {f^-}}-R^2\sqrt {f^+}( h_0^+)^{\prime} - \f {\xi_0} { \sqrt {f^-}^3} \right \}.
\end{align}


\end{document}